\documentclass[aps,preprint]{revtex4}%
\usepackage{amsfonts}
\usepackage{amsmath}
\usepackage{amssymb}
\usepackage{graphicx}%
\newtheorem{theorem}{Theorem}
\newtheorem{acknowledgement}[theorem]{Acknowledgement}

\begin{document}
\title{N-Dimensional non-abelian dilatonic, stable black holes and their Born-Infeld
extension }
\author{S. Habib Mazharimousavi$^{\ast}$}
\author{M. Halilsoy$^{\dag}$}
\author{Z. Amirabi$^{\ddag}$}
\affiliation{Department of Physics, Eastern Mediterranean University,}
\affiliation{G. Magusa, north Cyprus, Mersin-10, Turkey}
\affiliation{$^{\ast}$habib.mazhari@emu.edu.tr}
\affiliation{$^{\dagger}$mustafa.halilsoy@emu.edu.tr}
\affiliation{$^{\ddagger}$zahra.amirabi@emu.edu.tr}

\begin{abstract}
We find large classes of non-asymptotically flat Einstein-Yang-Mills-Dilaton
(EYMD) and Einstein-Yang-Mills-Born-Infeld-Dilaton (EYMBID) black holes in
N-dimensional spherically symmetric spacetime expressed in terms of the
quasilocal mass. Extension of the dilatonic YM solution to N-dimensions has
been possible by employing the generalized Wu-Yang ansatz. Another metric
ansatz, which aided in finding exact solutions is the functional dependence of
the radius function on the dilaton field. These classes of black holes are
stable against linear radial perturbations. In the limit of vanishing dilaton
we obtain Bertotti-Robinson (BR) type metrics with the topology of
$AdS_{2}\times S^{N-2}.$ Since connection can be established between dilaton
and a scalar field of Brans-Dicke (BD) type we obtain black hole solutions
also in the Brans-Dicke-Yang-Mills (BDYM) theory as well.

\end{abstract}
\maketitle

\section{Introduction}

Recently we have found black hole solutions in the Einstein-Yang-Mills (EYM)
theory by extending the Wu-Yang ansatz to higher dimensions \cite{1,2}. Both
the YM charge and the dimensionality of the space time played crucial roles to
determine the features of such black holes. It was found long ago, within the
context of Dirac monopole theory that Wu-Yang ansatz solves the static,
spherically symmetric YM equations for $N=4$ dimensional flat space
time\cite{3}. The $SO\left(  3\right)  $ gauge structure was derived from the
abelian electromagnetic (em) potential such that the internal and space time
indices were mixed together in the potential. By a similar analogy we extend
this idea to - nowadays fashionable$-N$ dimensional space times where
$SO\left(  N-1\right)  $ is obtained through a non-abelian gauge
transformation from the em potential within the static, spherically symmetric
metric ansatz. The YM gauge potential is chosen to depend only on the angular
variables and therefore they become independent of time ($t$) and the radial
coordinate ($r$). Upon this choice the YM potential becomes magnetic type and
by virtue of the metric ansatz the YM equations are easily satisfied. Such a
choice renders the duality principle to be automatically absent in the theory.
We note that by invoking the Birkhoff's theorem of general relativity $t$ and
$r$ can be interchanged appropriately in the metric, while the YM potential
preserves its form. The fact that the solutions obtained by this procedure
pertain to genuine non- abelian character is obvious from comparison with the
other known exact EYM solutions . The EYM\ solutions obtained by other
ansaetze\cite{4,5,6,7,8,9} constructed directly from the non-abelian character
and those obtained by our generalized Wu-Yang ansatz \cite{1,2} are the same.
We admit, however, that although our method yields exact solutions it is
restricted to spherical symmetry alone. Their solutions, on the other hand
\cite{4,5,6,7}, apply to less symmetric cases which at best can be expressed
in infinite series, and in certain limit, such as vanishing of a function,
they coincide with ours. Among other types, particle-like \cite{8} and
magnetic monopole \cite{9} solutions are discussed in even dimensions. To make
a comparison between EM and EYM solutions we refer to the different $r$ powers
in the solutions found so far. Specifically, the logarithmic term in the
metric for $N=5,$ EYM theory, for instance, is not encountered in the $N=5,$
EM theory \cite{1}. For $N=4$ it was verified on physical grounds that
although the metric remained unchanged, the geodesics particles felt the
non-abelian charges \cite{10}. We note that Ref. \cite{10} constitutes the
proper reference to be consulted in obtaining a YM solution from an EM
solution, which is stated as a theorem therein. Our study shows that the
distinction between the abelian and non-abelian contributions becomes more
transparent for $N>4.$ Let us\ note that throughout this paper by the
non-Abelian field we imply YM field whose higher dimensional version is
obtained by the generalized Wu-Yang ansatz.

It is well-known that in general relativity the field equations admit
solutions which, unlike the localized black holes can have different
properties. From this token we cite the cosmological solutions of de-Sitter
($dS$)/ Anti de-Sitter ($AdS$), the conformally flat and Bertoti-Robinson (BR)
type solutions \cite{11,12}, beside others in higher dimensions. In the EM
theory the conformally flat metric in $N=4$ is uniquely the BR\ metric whose
topology is $AdS_{2}\times S^{2}.$This extends to higher dimensions as
$AdS_{2}\times S^{N-2}$ which is no more conformally flat. The $N=4,$ BR
solution can be obtained from the extremal Reissner-Nordstrom (RN) black hole
solution through a limiting process. The latter represents a supersymmetric
soliton solution to connect different vacua of supergravity. For this reason
the BR geometry can be interpreted as a "throat" region between two
asymptotically flat space times. Also, since the source is pure homogenous
electromagnetic (em) field it is called an "em universe", which is free of
singularities. Its high degree of symmetry and singularity free properties
make BR space time attractive from both the string and supergravity theory
points of view . We recall that even for a satisfactory shell model
interpretation of an elementary particle, BR space time is proposed as a core
candidate \cite{13}. All these aspects ( and more), we believe, justifies to
make further studies on the BR\ space times, in particular for $N>4$, which
incorporates YM fields instead of the em fields.

In this paper we obtain new non-asymptotically flat dilatonic black hole
solutions and study their stability against linear perturbations\cite{14}.
Remarkably, they turn out to be stable against such perturbations. To obtain
such metrics we start with a general ansatz metric in the
Einstein-Yang-Mills-Dilaton (EYMD) theory. Our ansatz is of BR type instead of
the RN type so that in the limit of zero dilaton instead of \cite{1,2}, we
obtain BR type metrics. This leads us to a particular class of dilatonic
solutions coupled with the YM field. As expected, dilaton brings severe
restrictions on the space time which possesses singularities in general. In
the limit of zero dilaton we obtain a two parametric (i.e. Q and C) solution
which contains the BR solution as a subclass. The second parameter, which is
labeled as $C$, in a particular limit can be shown to correspond to the
quasilocal mass. Thus, keeping both $Q\neq0\neq C$ and a non-zero dilaton
gives us an asymptotically non-flat black hole model. The case $C=0$ (without
dilaton), yields a metric which is analogous to the BR metric\cite{15}.

Next, we extend our action to include the non-Abelian Born-Infeld (BI)
interaction which we phrase as Einstein-Yang-Mills-Born-Infeld-Dilaton
(EYMBID) theory. As it is well-known string / supergravity motivated
non-linear electrodynamics due to Born and Infeld \cite{16} received much
attention in recent years. Originally it was devised to eliminate divergences
due to point charges, which recovers the linear Maxwell's electrodynamics in a
particular limit (i.e. $\beta\rightarrow\infty$). Now it is believed that BI
action will provide significant contributions for the deep rooted problems of
quantum gravity. The BI action contains invariants in special combinations
under a square root term in analogy with the string theory Lagrangian. Since
our aim in this paper is to use non-Abelian fields instead of the em field we
shall employ the YM field which by our choice will be magnetic type. Some of
the solutions that we find for the EYMBID theory represent non-asymptotically
flat black holes. Unfortunately for an arbitrary dilatonic parameter the
solutions become untractable. One particular class of solutions on which we
shall elaborate will be again the BR type solutions for a vanishing dilaton.
We explore the possibility of finding conformally flat space time by choosing
particular BI parameter $\beta.$

After studying black holes in the dilatonic theory we proceed to establish
connection with the Brans-Dicke (BD) scalar field through a conformal
transformation and explore black holes in the latter as well. Coupling of BD
scalar field with YM field follows under the similar line of consideration.

The organization of the paper is as follows. In Sec. $II$ we introduce the
EYMD gravity, its field equations, their solutions and investigate their
stability. The Born-Infeld (BI) extension follows in Sec. $III$. Sec. $IV$
confines black holes in the Brans-Dicke-YM theory. The paper is completed with
conclusion in Sec. $V$.

\section{Field Equations and the metric ansatz for EYMD gravity}

The $N\left(  =n+1\right)  -$dimensional action in the EYMD theory is given by
$(G=1)$%
\begin{gather}
I=-\frac{1}{16\pi}\int\nolimits_{\mathcal{M}}d^{n+1}x\sqrt{-g}\left(
R-\frac{4}{n-1}\left(  \mathbf{\nabla}\Phi\right)  ^{2}+\mathcal{L}\left(
\Phi\right)  \right)  -\frac{1}{8\pi}\int\nolimits_{\partial\mathcal{M}}%
d^{n}x\sqrt{-h}K,\\
\mathcal{L}\left(  \Phi\right)  =-e^{-4\alpha\Phi/\left(  n-1\right)
}\mathbf{Tr}(F_{\lambda\sigma}^{\left(  a\right)  }F^{\left(  a\right)
\lambda\sigma}),\nonumber
\end{gather}
where%
\begin{equation}
\mathbf{Tr}(.)=\overset{\left(  n\right)  (n-1)/2}{\underset{a=1}{%
{\textstyle\sum}
}\left(  .\right)  ,}%
\end{equation}
$\Phi$ refers to the dilaton scalar potential (we should comment that in this
work we are interested in a spherical symmetric dilatonic potential, i.e.
$\Phi=\Phi\left(  r\right)  $) and $\alpha$ denotes the dilaton parameter
while the second term is the surface integral with its induced metric $h_{ij}$
and trace $K$ of its extrinsic curvature. Herein $R$ is the usual Ricci scalar
and $\mathbf{F}^{\left(  a\right)  }=F_{\mu\nu}^{\left(  a\right)  }dx^{\mu
}\wedge dx^{\nu}$ are the YM field $2-$forms (with $\wedge$ indicating the
wedge product) which are given by \cite{1,2}%
\begin{equation}
\mathbf{F}^{\left(  a\right)  }=\mathbf{dA}^{\left(  a\right)  }+\frac
{1}{2\sigma}C_{\left(  b\right)  \left(  c\right)  }^{\left(  a\right)
}\mathbf{A}^{\left(  b\right)  }\wedge\mathbf{A}^{\left(  c\right)  }%
\end{equation}
with structure constants $C_{\left(  b\right)  \left(  c\right)  }^{\left(
a\right)  }$ (see Appendix A) while $\sigma$ is a coupling constant and
$\mathbf{A}^{\left(  a\right)  }=A_{\mu}^{\left(  a\right)  }dx^{\mu}$ are the
potential $1-$forms. Our choice of YM potential $\mathbf{A}^{\left(  a\right)
}$ follows from the higher dimensional Wu-Yang ansatz \cite{1,2} where
$\sigma$ is expressed in terms of the YM charge. Variations of the action with
respect to the gravitational field $g_{\mu\nu}$ and the scalar field $\Phi$
lead, respectively to the EYMD field equations
\begin{gather}
R_{\mu\nu}=\frac{4}{n-1}\mathbf{\partial}_{\mu}\Phi\partial_{\nu}%
\Phi+2e^{-4\alpha\Phi/\left(  n-1\right)  }\left[  \mathbf{Tr}\left(
F_{\mu\lambda}^{\left(  a\right)  }F_{\nu}^{\left(  a\right)  \ \lambda
}\right)  -\frac{1}{2\left(  n-1\right)  }\mathbf{Tr}\left(  F_{\lambda\sigma
}^{\left(  a\right)  }F^{\left(  a\right)  \lambda\sigma}\right)  g_{\mu\nu
}\right]  ,\\
\nabla^{2}\Phi=-\frac{1}{2}\alpha e^{-4\alpha\Phi/\left(  n-1\right)
}\mathbf{Tr}(F_{\lambda\sigma}^{\left(  a\right)  }F^{\left(  a\right)
\lambda\sigma}),
\end{gather}
where $R_{\mu\nu}$ is the Ricci tensor. Variation with respect to the gauge
potentials $\mathbf{A}^{\left(  a\right)  }$ yields the YM equations%

\begin{equation}
\mathbf{d}\left(  e^{-4\alpha\Phi/\left(  n-1\right)  \star}\mathbf{F}%
^{\left(  a\right)  }\right)  +\frac{1}{\sigma}C_{\left(  b\right)  \left(
c\right)  }^{\left(  a\right)  }e^{-4\alpha\Phi/\left(  n-1\right)
}\mathbf{A}^{\left(  b\right)  }\wedge^{\star}\mathbf{F}^{\left(  c\right)
}=0
\end{equation}
in which the hodge star $^{\star}$ means duality. In the next section we shall
present solutions to the foregoing equations in N-dimension. Wherever it is
necessary we shall supplement our discussion by resorting to the particular
case $N=5$. Let us remark that for $N=4$ case since the YM field becomes gauge
equivalent to the em field the metrics are still of RN/BR, therefore we shall
ignore the case $N=4$.

\subsection{N-dimensional solution}

In $N\left(  =n+1\right)  -$dimensions, we choose a spherically symmetric
metric ansatz
\begin{equation}
ds^{2}=-f\left(  r\right)  dt^{2}+\frac{dr^{2}}{f\left(  r\right)  }+h\left(
r\right)  ^{2}d\Omega_{n-1}^{2},
\end{equation}
where%
\begin{equation}
d\Omega_{n-1}^{2}=d\theta_{1}^{2}+\underset{i=2}{\overset{n-2}{%
{\textstyle\sum}
}}\underset{j=1}{\overset{i-1}{%
{\textstyle\prod}
}}\sin^{2}\theta_{j}\;d\theta_{i}^{2},\text{ \ \ }0\leq\theta_{n-1}\leq
2\pi,\text{ }0\leq\theta_{k\neq n-1}\leq\pi.
\end{equation}
while $f\left(  r\right)  $ and $h\left(  r\right)  $ are two functions to be
determined. Our gauge potential ansatz is \cite{1,2}%
\begin{align}
\mathbf{A}^{(a)}  &  =\frac{Q}{r^{2}}C_{\left(  i\right)  \left(  j\right)
}^{\left(  a\right)  }\ x^{i}dx^{j},\text{ \ \ }Q=\text{YM magnetic charge,
\ }r^{2}=\overset{n}{\underset{i=1}{\sum}}x_{i}^{2},\\
2  &  \leq j+1\leq i\leq n,\text{ \ and \ }1\leq a\leq n(n-1)/2,\nonumber\\
x_{1}  &  =r\cos\theta_{n-1}\sin\theta_{n-2}...\sin\theta_{1},\text{ }%
x_{2}=r\sin\theta_{n-1}\sin\theta_{n-2}...\sin\theta_{1},\nonumber\\
\text{ }x_{3}  &  =r\cos\theta_{n-2}\sin\theta_{n-3}...\sin\theta_{1},\text{
}x_{4}=r\sin\theta_{n-2}\sin\theta_{n-3}...\sin\theta_{1},\nonumber\\
&  ...\nonumber\\
x_{n}  &  =r\cos\theta_{1}.\nonumber
\end{align}
We note that the structure constant $C_{ij}^{a}$ are found similar to the case
$N=5$ as described in Appendix A. The YM equations (6) are satisfied and the
field equations become%
\begin{align}
\nabla^{2}\Phi &  =-\frac{1}{2}\alpha e^{-4\alpha\Phi/\left(  n-1\right)
}\mathbf{Tr}(F_{\lambda\sigma}^{\left(  a\right)  }F^{\left(  a\right)
\lambda\sigma})\\
R_{tt}  &  =\frac{e^{-4\alpha\Phi/\left(  n-1\right)  }f}{\left(  n-1\right)
}\mathbf{Tr}(F_{\lambda\sigma}^{\left(  a\right)  }F^{\left(  a\right)
\lambda\sigma})\\
R_{rr}  &  =\frac{4\left(  \Phi^{\prime}\right)  ^{2}}{\left(  n-1\right)
}-\frac{e^{-4\alpha\Phi/\left(  n-1\right)  }}{\left(  n-1\right)
f}\mathbf{Tr}(F_{\lambda\sigma}^{\left(  a\right)  }F^{\left(  a\right)
\lambda\sigma})\\
R_{\theta_{i}\theta_{i}}  &  =\frac{2\left(  n-2\right)  Q^{2}e^{-4\alpha
\Phi/\left(  n-1\right)  }}{h^{2}}-\frac{h^{2}e^{-4\alpha\Phi/\left(
n-1\right)  }}{\left(  n-1\right)  }\mathbf{Tr}(F_{\lambda\sigma}^{\left(
a\right)  }F^{\left(  a\right)  \lambda\sigma}),
\end{align}
in which we note that the remaining angular Ricci parts add no new conditions.
A proper ansatz for $h\left(  r\right)  $ now is%
\begin{align}
h\left(  r\right)   &  =Ae^{-2\alpha\Phi/\left(  n-1\right)  }\\
\text{(}A  &  =\text{constant)}\nonumber
\end{align}
which, after knowing
\begin{equation}
\mathbf{Tr}(F_{\lambda\sigma}^{\left(  a\right)  }F^{\left(  a\right)
\lambda\sigma})=\frac{\left(  n-1\right)  \left(  n-2\right)  Q^{2}}{h^{4}}%
\end{equation}
and eliminating $f\left(  r\right)  $ from Eq.s (11) and (12) one gets
\begin{equation}
\Phi=-\frac{\left(  n-1\right)  }{2}\frac{\alpha\ln r}{\alpha^{2}+1}.
\end{equation}
Upon substitution of $\Phi$ and $h\left(  r\right)  $ into the Eq.s (10)-(13)
we get three new equations
\begin{gather}
\left(  n-1\right)  \left[  r\left(  \alpha^{2}+1\right)  f^{\prime}+\left(
\left(  n-2\right)  \alpha^{2}-1\right)  f\right]  -\left(  \frac{\left(
n-1\right)  \left(  n-2\right)  Q^{2}}{A^{4}}\right)  \left(  \alpha
^{2}+1\right)  ^{2}r^{\left(  \frac{2}{\alpha^{2}+1}\right)  }=0\\
\left(  n-1\right)  \left[  r\left(  \alpha^{2}+1\right)  f^{\prime\prime
}+\left(  n-1\right)  \alpha^{2}f^{\prime}\right]  -2\left(  \frac{\left(
n-1\right)  \left(  n-2\right)  Q^{2}}{A^{4}}\right)  \left(  \alpha
^{2}+1\right)  r^{\left(  -\frac{\alpha^{2}-1}{\alpha^{2}+1}\right)  }=0\\
\left(  \alpha^{2}+1\right)  ^{2}\left(  n-2\right)  \left(  Q^{2}%
-A^{2}\right)  r^{2}+\nonumber\\
A^{4}\alpha^{2}\left(  \alpha^{2}+1\right)  f^{\prime}r^{\left(  \frac
{3\alpha^{2}+1}{\alpha^{2}+1}\right)  }+\alpha^{2}\left(  \left(  n-2\right)
\alpha^{2}-1\right)  A^{4}fr^{\left(  \frac{2\alpha^{2}}{\alpha^{2}+1}\right)
}=0.
\end{gather}
Eq. (17) yields the integral for $f\left(  r\right)  $%
\begin{align}
f\left(  r\right)   &  =\Xi\left(  1-\left(  \frac{r_{+}}{r}\right)
^{\frac{\left(  n-2\right)  \alpha^{2}+1}{\alpha^{2}+1}}\right)  r^{\frac
{2}{\alpha^{2}+1}},\\
\Xi &  =\frac{\left(  n-2\right)  }{\left(  \left(  n-2\right)  \alpha
^{2}+1\right)  Q^{2}}%
\end{align}
and the equations (18) and (19) imply that $A$ must satisfy the following
constraint%
\begin{equation}
A^{2}=Q^{2}\left(  \alpha^{2}+1\right)  .
\end{equation}
\ One may notice that, with the solution (20), (7) becomes a
non-asymptotically flat metric and therefore the ADM mass can not be defined.
Following the quasilocal mass formalism introduced by Brown and York \cite{17}
it is known that, a spherically symmetric N-dimensional metric solution as
\begin{equation}
ds^{2}=-F\left(  R\right)  ^{2}dt^{2}+\frac{dR^{2}}{G\left(  R\right)  ^{2}%
}+R^{2}d\Omega_{N-2}^{2},
\end{equation}
admits a quasilocal mass $M_{QL}$ defined by \cite{18,19}%
\begin{equation}
M_{QL}=\frac{N-2}{2}R_{B}^{N-3}F\left(  R_{B}\right)  \left(  G_{ref}\left(
R_{B}\right)  -G\left(  R_{B}\right)  \right)  .
\end{equation}
Here $G_{ref}\left(  R\right)  $ is an arbitrary reference function, which
guarantees having zero quasilocal mass once the matter source is turned off
and $R_{B}$ is the radius of the spacelike hypersurface boundary. Applying
this formalism to the solution (20), one obtains the horizon $r_{+}$ in terms
of $M_{QL}$ as
\begin{equation}
r_{+}=\left(  \frac{4\left(  \alpha^{2}+1\right)  M_{QL}}{\left(  n-1\right)
\Xi\alpha^{2}A^{n-1}}\right)  .
\end{equation}
Having the radius of horizon, one may use the usual definition of the Hawking
temperature to calculate
\begin{equation}
T_{H}=\frac{1}{4\pi}\left\vert f^{\prime}\left(  r_{+}\right)  \right\vert
=\frac{\Xi}{4\pi}\frac{\left[  \left(  n-2\right)  \alpha^{2}+1\right]
}{\left(  \alpha^{2}+1\right)  }\left(  r_{+}\right)  ^{\gamma}%
\end{equation}
where $\Xi$ and $r_{+}$ are given above and $\gamma=\frac{1-\alpha^{2}%
}{1+\alpha^{2}}$.

In order to see the singularity of the spacetime we calculate the scalar
invariants, which are tedious for general N, for this reason we restrict
ourselves to the case $N=5$ alone. The scalar invariants for $N=5$ are as
follows
\begin{align}
R  &  =\frac{\omega_{1}}{r^{\frac{4\alpha^{2}+1}{\alpha^{2}+1}}}+\frac
{\sigma_{1}}{r^{\frac{2\alpha^{2}}{\alpha^{2}+1}}},\\
R_{\mu\nu}R^{\mu\nu}  &  =\frac{\omega_{2}}{r^{\frac{6\alpha^{2}+1}{\alpha
^{2}+1}}}+\frac{\omega_{3}}{r^{2\frac{4\alpha^{2}+1}{\alpha^{2}+1}}}%
+\frac{\sigma_{2}}{r^{\frac{4\alpha^{2}}{\alpha^{2}+1}}},\\
R_{\mu\nu\alpha\beta}R^{\mu\nu\alpha\beta}  &  =\frac{\omega_{4}}%
{r^{\frac{6\alpha^{2}+1}{\alpha^{2}+1}}}+\frac{\omega_{5}}{r^{2\frac
{4\alpha^{2}+1}{\alpha^{2}+1}}}+\frac{\sigma_{3}}{r^{\frac{4\alpha^{2}}%
{\alpha^{2}+1}}}%
\end{align}
where $\omega_{i}$ and $\sigma_{i}$ are some constants and
\begin{align}
\underset{\alpha\rightarrow0}{\lim}\omega_{i}  &  =0,\text{ \ \ }%
\underset{\alpha\rightarrow0}{\lim}\sigma_{1}=\frac{2}{Q^{2}},\\
\text{\ \ }\underset{\alpha\rightarrow0}{\lim}\sigma_{2}  &  =\frac{20}{Q^{4}%
},\text{ \ \ }\underset{\alpha\rightarrow0}{\lim}\sigma_{3}=\frac{33}{Q^{4}%
}.\text{ }\nonumber
\end{align}
These results show that, for non-zero dilaton field (i.e. $\alpha\neq0$), the
origin is singular whereas for $\alpha=0$ (as a limit), we have a regular
spacetime. Although these results have been found for $N=5$, it is our belief
that for a general $N>5$ these behaviors do not show much difference.

\subsubsection{Linear dilaton}

Setting $\alpha=1,$ gives the linear dilaton solution (20) as%
\begin{align}
f\left(  r\right)   &  =\frac{\left(  n-2\right)  }{\left(  n-1\right)  Q^{2}%
}\left(  1-\left(  \frac{r_{+}}{r}\right)  ^{\frac{n-1}{2}}\right)  r,\text{
\ \ }h\left(  r\right)  ^{2}=2Q^{2}r\\
\text{\ \ }r_{+}  &  =\left(  \frac{2^{\frac{7-n}{2}}M_{QL}}{\left(
n-2\right)  \left(  \left\vert Q\right\vert \right)  ^{n-3}}\right)  .
\end{align}
One can use the standard\ way to find the high frequency limit of Hawking
temperature at the horizon, which means that%
\begin{equation}
T_{H}=\frac{1}{4\pi}\left\vert f^{\prime}\left(  r_{+}\right)  \right\vert
=\frac{\left(  n-2\right)  }{8\pi Q^{2}}.
\end{equation}
Furthermore, $M_{QL}$ is an integration constant which is identified as
quasilocal mass, so one may set this constant to be zero to get the line
element
\begin{align}
ds^{2}  &  =-\Xi rdt^{2}+\frac{dr^{2}}{\Xi r}+2Q^{2}rd\Omega_{n-1}^{2},\\
\Xi &  =\frac{\left(  n-2\right)  }{\left(  n-1\right)  Q^{2}}.
\end{align}
By a simple transformation $r=e^{\Xi\rho}$ this line element transforms into%
\begin{equation}
ds^{2}=\Xi e^{\Xi\rho}\left(  -dt^{2}+d\rho^{2}+\frac{2\left(  n-1\right)
Q^{4}}{\left(  n-2\right)  }d\Omega_{n-1}^{2}\right)
\end{equation}
which represents a conformal $M_{2}\times S_{n-1}$ space time with the radius
of $S_{n-1}$ equal to $\sqrt{\frac{2\left(  n-1\right)  }{\left(  n-2\right)
}}Q^{2}.$

\subsubsection{BR limit of the solution}

In the zero dilaton limit $\alpha=0,$ we express our metric function in the
form of%
\begin{align}
f\left(  r\right)   &  =\Xi_{\circ}\left(  r-r_{+}\right)  r,\text{ \ }%
\Xi_{\circ}=\frac{\left(  n-2\right)  }{Q^{2}},\\
h^{2}  &  =A_{\circ}^{2}=Q^{2}.
\end{align}
In $N(=n+1)-$dimensions we also set $r_{+}=0,$ $r=\frac{1}{\rho}$ and
$\tau=\Xi_{\circ}t,$ to transform the metric (7) into%
\begin{equation}
ds^{2}=\frac{Q^{2}}{\left(  n-2\right)  }\left(  \frac{-d\tau^{2}+d\rho^{2}%
}{\rho^{2}}+\left(  n-2\right)  d\Omega_{n-1}^{2}\right)  .
\end{equation}
This is in the BR form with the topological structure $AdS_{2}\times S^{n-1},$
where the radius of the $S^{n-1}$ sphere is $\sqrt{n-2}.$

\subsubsection{$AdS_{2}\times S^{N-2}$ topology for $0<\alpha<1$}

In this section we shall show that, the general solution given in Eq. (20),
for some specific values for $0<\alpha<1,$ may also represent a conformally
flat space time. To this end, we set $r_{+}=0,$ and apply the following
transformation
\begin{align}
r  &  =\left(  \Xi\frac{1-\alpha^{2}}{1+\alpha^{2}}\rho\right)  ^{-\frac
{1+\alpha^{2}}{1-\alpha^{2}}},\\
\Xi &  =\frac{\left(  n-2\right)  }{\left(  \left(  n-2\right)  \alpha
^{2}+1\right)  Q^{2}},
\end{align}
to get%
\begin{equation}
ds^{2}=\left(  \Xi\right)  ^{-\frac{1+\alpha^{2}}{1-\alpha^{2}}}\left(
\frac{1-\alpha^{2}}{1+\alpha^{2}}\right)  ^{-\frac{2}{1-\alpha^{2}}}%
\rho^{-\frac{2\alpha^{2}}{1-\alpha^{2}}}\left(  \frac{-d\tau^{2}+d\rho^{2}%
}{\rho^{2}}+\Xi A^{2}\left(  \frac{1-\alpha^{2}}{1+\alpha^{2}}\right)
^{2}d\Omega_{n-1}^{2}\right)  .
\end{equation}
To have a conformally flat space time, we impose $\Xi A^{2}\left(
\frac{1-\alpha^{2}}{1+\alpha^{2}}\right)  ^{2}$ to be one, i.e.%
\begin{equation}
\frac{\left(  n-2\right)  \left(  1-\alpha^{2}\right)  ^{2}}{\left(  \left(
n-2\right)  \alpha^{2}+1\right)  \left(  \alpha^{2}+1\right)  }=1
\end{equation}
and therefore yields, $\alpha^{2}=\frac{n-3}{3n-5}.$ The line element (42)
takes the form of a conformally flat space time, namely%
\begin{align}
ds^{2}  &  =a\left(  \rho\right)  \left(  \frac{-d\tau^{2}+d\rho^{2}}{\rho
^{2}}+d\Omega_{n-1}^{2}\right)  ,\\
a\left(  \rho\right)   &  =2^{\frac{3n-5}{n-1}}\left(  n-2\right)  \left(
\frac{Q^{2}}{3n-5}\right)  ^{2\frac{n-2}{n-1}}\left(  n-1\right)  ^{\frac
{n-3}{n-1}}\rho^{-\frac{2\alpha^{2}}{1-\alpha^{2}}}.
\end{align}

\subsection{Linear Stability of the EYMD black holes}

In this chapter we follow a similar method used by Yazadjiev \cite{14} to
investigate the stability of the possible EYMD black hole solutions,
introduced previously, in terms of a linear radial perturbation. Although this
method is applicable to any dimensions we confine ourselves to the
five-dimensional black hole case given by Eq. (7). To do so we assume that our
dilatonic scalar field $\Phi\left(  r\right)  $ changes into $\Phi\left(
r\right)  +\psi\left(  t,r\right)  ,$ in which $\psi\left(  t,r\right)  $ is
very weak compared to the original dilaton field and we call it the perturbed
term. As a result we choose our perturbed metric as
\begin{equation}
ds^{2}=-f\left(  r\right)  e^{\Gamma\left(  t,r\right)  }dt^{2}+e^{\chi\left(
t,r\right)  }\frac{dr^{2}}{f\left(  r\right)  }+h\left(  r\right)  ^{2}%
d\Omega_{3}^{2}.
\end{equation}
One should notice that, since our gauge potentials are magnetic, the YM
equations (6) are satisfied. The linearized version of the field equations
(10-13) plus one extra term of $R_{tr}$ are given by%
\begin{gather}
R_{tr}:\frac{3}{2}\frac{\chi_{t}\left(  t,r\right)  h^{\prime}\left(
r\right)  }{h\left(  r\right)  }=\frac{4}{3}\partial_{r}\Phi\left(  r\right)
\partial_{t}\psi\left(  t,r\right) \\
\nabla_{\circ}^{2}\psi-\chi\nabla_{\circ}^{2}\Phi+\frac{1}{2}\left(
\Gamma-\chi\right)  _{r}\Phi^{\prime}f=\frac{4\alpha^{2}e^{\frac{4}{3}%
\alpha\Phi}}{Q^{2}\left(  \alpha^{2}+1\right)  ^{2}}\psi\\
R_{\theta\theta}:\left(  2-R_{\circ\theta\theta}\right)  \chi-\frac{1}%
{2}hh^{\prime}f\left(  \Gamma-\chi\right)  _{r}=\frac{8\alpha}{3\left(
\alpha^{2}+1\right)  }\psi
\end{gather}
in which a lower index $_{\circ}$ represents the quantity in the unperturbed
metric. First equation in this set implies%
\begin{equation}
\chi\left(  t,r\right)  =-\frac{4}{3\alpha}\psi\left(  t,r\right)
\end{equation}
which after making substitutions in the two latter equations and eliminating
the $\left(  \Gamma-\chi\right)  _{r}$ one finds%
\begin{equation}
\nabla_{\circ}^{2}\psi\left(  t,r\right)  -U\left(  r\right)  \psi\left(
t,r\right)  =0
\end{equation}
where%
\begin{equation}
U\left(  r\right)  =\frac{4e^{\frac{4}{3}\alpha\Phi}}{Q^{2}\left(
1+\alpha^{2}\right)  }=\frac{4}{Q^{2}\left(  1+\alpha^{2}\right)
r^{\frac{2\alpha^{2}}{1+\alpha^{2}}}}.
\end{equation}
To get these results we have implicitly used the constraint (22) on A. Again
by imposing the same constraint , one can show that $U\left(  r\right)  $ is
positive. It is not difficult to apply the separation method on (51) to get
\begin{equation}
\psi\left(  t,r\right)  =e^{\pm\epsilon t}\zeta\left(  r\right)  ,\text{
\ \ }\nabla_{\circ}^{2}\zeta\left(  r\right)  -U_{eff}\left(  r\right)
\zeta\left(  r\right)  =0,\text{ \ \ }U_{eff}\left(  r\right)  =\left(
\frac{\epsilon^{2}}{f}+U\left(  r\right)  \right)  ,
\end{equation}
where $\epsilon$ is a constant. Since $U_{eff}\left(  r\right)  $ is positive
one can easily show that, for any real value for $\epsilon$ there exists a
solution for $\zeta\left(  r\right)  $ which is not bounded. In other words by
the linear perturbation our black hole solution is stable for any value of
$\epsilon.$ As a limit of this proof, one may set $\alpha=0,$ which recovers
the BR case.

We remark that with little addition this method can be easily extended to any
higher dimensions. This implies that the N-dimensional EYMD black holes are
stable under the linear perturbation.

\section{Field equations and the metric ansatz for EYMBID gravity}

The $N\left(  =n+1\right)  -$dimensional action in the EYMBI-D theory is given
by $(G=1)$%
\begin{gather}
I=-\frac{1}{16\pi}\int\nolimits_{\mathcal{M}}d^{n+1}x\sqrt{-g}\left(
R-\frac{4}{n-1}\left(  \mathbf{\nabla}\Phi\right)  ^{2}+L\left(
\mathbf{F},\Phi\right)  \right)  -\frac{1}{8\pi}\int\nolimits_{\partial
\mathcal{M}}d^{n}x\sqrt{-h}K,\\
L\left(  \mathbf{F},\Phi\right)  =4\beta^{2}e^{4\alpha\Phi/\left(  n-1\right)
}\left(  1-\sqrt{1+\frac{\mathbf{Tr}(F_{\lambda\sigma}^{\left(  a\right)
}F^{\left(  a\right)  \lambda\sigma})e^{-8\alpha\Phi/\left(  n-1\right)  }%
}{2\beta^{2}}}\right)  =\\
4\beta^{2}e^{4\alpha\Phi/\left(  n-1\right)  }\mathcal{L}\left(  X\right)
,\nonumber
\end{gather}
where%
\begin{equation}
\mathcal{L}\left(  X\right)  =1-\sqrt{1+X},\text{ \ \ }X=\frac{\mathbf{Tr}%
(F_{\lambda\sigma}^{\left(  a\right)  }F^{\left(  a\right)  \lambda\sigma
})e^{-8\alpha\Phi/\left(  n-1\right)  }}{2\beta^{2}},\text{ \ \ }%
\mathbf{Tr}(.)=\overset{n(n-1)/2}{\underset{a=1}{%
{\textstyle\sum}
}\left(  .\right)  ,}%
\end{equation}
while the rest of the parameters are defined as before. Variations of the
EYMBID action with respect to the gravitational field $g_{\mu\nu}$ and the
scalar field $\Phi$ lead respectively to the correspondence EYMBID field
equations
\begin{align}
R_{\mu\nu}  &  =\frac{4}{n-1}\mathbf{\partial}_{\mu}\Phi\partial_{\nu}%
\Phi-4e^{-4\alpha\Phi/\left(  n-1\right)  }\left(  \mathbf{Tr}\left(
F_{\mu\lambda}^{\left(  a\right)  }F_{\nu}^{\left(  a\right)  \ \lambda
}\right)  \partial_{X}\mathcal{L}\left(  X\right)  \right)  +\\
&  \frac{4\beta^{2}}{n-1}e^{4\alpha\Phi/\left(  n-1\right)  }\mathcal{K}%
\left(  X\right)  g_{\mu\nu},\nonumber\\
\nabla^{2}\Phi &  =2\alpha\beta^{2}e^{4\alpha\Phi/\left(  n-1\right)
}\mathcal{K}\left(  X\right)  ,\text{\ \ \ \ \ }%
\end{align}
where we have abbreviated%
\begin{align}
\mathcal{K}\left(  X\right)   &  =2X\partial_{X}\mathcal{L}\left(  X\right)
-\mathcal{L}\left(  X\right) \\
(\partial_{X}\mathcal{L}\left(  X\right)   &  =-\frac{1}{\sqrt{1+X}%
})\text{.\ }\nonumber
\end{align}
Variation with respect to the gauge potentials $\mathbf{A}^{\left(  a\right)
}$ yields the new relevant YM equations%

\begin{equation}
\mathbf{d}\left(  e^{-4\alpha\Phi/\left(  n-1\right)  \star}\mathbf{F}%
^{\left(  a\right)  }\partial_{X}\mathcal{L}\left(  X\right)  \right)
+\frac{1}{\sigma}C_{\left(  b\right)  \left(  c\right)  }^{\left(  a\right)
}e^{-4\alpha\Phi/\left(  n-1\right)  }\partial_{X}\mathcal{L}\left(  X\right)
\mathbf{A}^{\left(  b\right)  }\wedge^{\star}\mathbf{F}^{\left(  c\right)
}=0.
\end{equation}
It is remarkable to observe that the field equations (57-59) in the limit of
$\beta\rightarrow\infty,$ reduce to the Eq.s (4-6), which are the field
equations for the EYMD theory. Also in the limit of $\beta\rightarrow0$, Eq.s
(57-59) give%
\begin{align}
R_{\mu\nu}  &  =\frac{4}{n-1}\mathbf{\partial}_{\mu}\Phi\partial_{\nu}\Phi,\\
\nabla^{2}\Phi &  =0\text{\ }%
\end{align}
which refer to the gravity coupled with a massless scalar field.

\subsection{N-dimensional solution}

In $N\left(  =n+1\right)  -$dimensions, we again, adopt the metric ansatz (7)
and our YM potentials are given by Eq. (9). N-dimensional YM equations (60)
are satisfied while the field equations imply the following set of four
equations%
\begin{align}
\nabla^{2}\Phi &  =2\alpha\beta^{2}e^{4\alpha\Phi/\left(  n-1\right)
}\mathcal{K}\left(  X\right) \\
R_{tt}  &  =-\frac{4\beta^{2}e^{4\alpha\Phi/\left(  n-1\right)  }f}{\left(
n-1\right)  }\mathcal{K}\left(  X\right) \\
R_{rr}  &  =\frac{4\left(  \Phi^{\prime}\right)  ^{2}}{\left(  n-1\right)
}+\frac{4\beta^{2}e^{4\alpha\Phi/\left(  n-1\right)  }}{\left(  n-1\right)
f}\mathcal{K}\left(  X\right) \\
R_{\theta_{i}\theta_{i}}  &  =\frac{-4\left(  n-2\right)  Q^{2}e^{-4\alpha
\Phi/\left(  n-1\right)  }}{h^{2}}\partial_{X}\mathcal{L}+\frac{4h^{2}%
\beta^{2}e^{4\alpha\Phi/\left(  n-1\right)  }}{\left(  n-1\right)
}\mathcal{K}\left(  X\right)  .
\end{align}
in which $X$ is defined by (56). We use the same ansatz for $h\left(
r\right)  $ as Eq. (14)which gives%
\begin{equation}
X=\frac{\left(  n-1\right)  \left(  n-2\right)  Q^{2}}{2\beta^{2}A^{4}}%
\end{equation}
and therefore, after eliminating $f\left(  r\right)  $ from Eq.s (64) and
(65), leads to (16). Upon substitution of $\Phi$ and $h\left(  r\right)  $
into the Eq.s (63)-(66) we find the following equations
\begin{gather}
\left(  n-1\right)  \left[  r\left(  \alpha^{2}+1\right)  f^{\prime}+\left(
\left(  n-2\right)  \alpha^{2}-1\right)  f\right]  +4\beta^{2}\mathcal{K}%
\left(  X\right)  \left(  \alpha^{2}+1\right)  ^{2}r^{\left(  \frac{2}%
{\alpha^{2}+1}\right)  }=0\\
\left(  n-1\right)  \left[  r\left(  \alpha^{2}+1\right)  f^{\prime\prime
}+\left(  n-1\right)  \alpha^{2}f^{\prime}\right]  +8\beta^{2}\mathcal{K}%
\left(  X\right)  \left(  \alpha^{2}+1\right)  r^{\left(  -\frac{\alpha^{2}%
-1}{\alpha^{2}+1}\right)  }=0\\
\left(  \alpha^{2}+1\right)  ^{2}\left(  4\beta^{2}A^{4}\mathcal{K}\left(
X\right)  -(4Q^{2}\partial_{X}\mathcal{L}+A^{2})\left(  n-1\right)  \left(
n-2\right)  \right)  r^{2}+\\
\left(  n-1\right)  A^{4}\alpha^{2}\left(  \alpha^{2}+1\right)  f^{\prime
}r^{\left(  \frac{3\alpha^{2}+1}{\alpha^{2}+1}\right)  }+\left(  n-1\right)
\alpha^{2}\left(  \left(  n-2\right)  \alpha^{2}-1\right)  A^{4}fr^{\left(
\frac{2\alpha^{2}}{\alpha^{2}+1}\right)  }=0.\nonumber
\end{gather}
Eq. (68) yields the integral for $f\left(  r\right)  $%
\begin{align}
f\left(  r\right)   &  =\Xi\left(  1-\left(  \frac{r_{+}}{r}\right)
^{\frac{\left(  n-2\right)  \alpha^{2}+1}{\alpha^{2}+1}}\right)  r^{\frac
{2}{\alpha^{2}+1}},\\
\Xi &  =-\frac{4\beta^{2}\left(  \alpha^{2}+1\right)  ^{2}\mathcal{K}\left(
X\right)  }{\left(  n-1\right)  \left(  \left(  n-2\right)  \alpha
^{2}+1\right)  }%
\end{align}
in which $r_{+}$ is an integration constant connected to the quasi local mass
i.e.,%
\begin{equation}
r_{+}=\left(  \frac{4\left(  \alpha^{2}+1\right)  M_{QL}}{\left(  n-1\right)
\Xi\alpha^{2}A^{n-1}}\right)
\end{equation}
and $\mathcal{K}\left(  X\right)  $ is abbreviated as in (59). This solution
satisfies Eq. (69), but from Eq. (70) $A$ must satisfy the constraint%
\begin{equation}
4\mathcal{K}\left(  X\right)  \beta^{2}A^{4}\left(  \alpha^{2}-1\right)
+\left(  n-1\right)  \left(  n-2\right)  \left(  4Q^{2}\partial_{X}%
\mathcal{L}+A^{2}\right)  =0.
\end{equation}

\subsubsection{Linear dilaton}

In the linear dilaton case i.e., $\alpha=1,$ Eq. (71) yields%
\begin{equation}
f\left(  r\right)  =\Xi\left(  1-\left(  \frac{r_{+}}{r}\right)
^{\frac{\left(  n-2\right)  +1}{2}}\right)  r,\text{ \ \ }h\left(  r\right)
=A\sqrt{r}\text{, \ \ }r_{+}=\left(  \frac{8M_{QL}}{\left(  n-1\right)  \Xi
A^{n-1}}\right)
\end{equation}
in which
\begin{equation}
A^{2}=2Q^{2}\sqrt{1-\frac{Q_{cri}^{2}}{Q^{2}}},\text{ \ \ }\Xi=\frac{2\left(
n-2\right)  }{\left(  n-1\right)  Q_{cri}^{2}}\left(  1-\sqrt{1-\frac
{Q_{cri}^{2}}{Q^{2}}}\right)
\end{equation}
where
\begin{equation}
Q_{cri}^{2}=\frac{\left(  n-1\right)  \left(  n-2\right)  }{8\beta^{2}}%
\end{equation}
and $Q^{2}\geq Q_{cri}^{2}.$

In this case one may set $\Xi=A=1$ to get
\begin{equation}
ds^{2}=-\left(  1-\left(  \frac{r_{+}}{r}\right)  ^{\frac{\left(  n-2\right)
+1}{2}}\right)  rdt^{2}+\frac{1}{\left(  1-\left(  \frac{r_{+}}{r}\right)
^{\frac{\left(  n-2\right)  +1}{2}}\right)  r}dr^{2}+rd\Omega_{n-1}^{2}.
\end{equation}

\subsubsection{BR limit of the solution}

In the zero dilaton limit $\alpha=0,$ we express our metric functions (71) in
the form
\begin{align}
f\left(  r\right)   &  =\Xi_{\circ}\left(  r-r_{+}\right)  r,\text{ \ }%
\Xi_{\circ}=\frac{8\beta^{2}\left(  n-2\right)  }{\left(  n-1\right)  \left(
n-2\right)  +8\beta^{2}Q^{2}},\\
h^{2}  &  =A_{\circ}^{2}=Q^{2}-\frac{\left(  n-1\right)  \left(  n-2\right)
}{8\beta^{2}}.
\end{align}
In $N(=n+1)-$dimensions we also set $r_{+}=0,$ $r=\frac{1}{\rho}$ and
$\tau=\Xi_{\circ}t,$ to transform the metric (7) into%
\begin{equation}
ds^{2}=\frac{1}{\Xi_{\circ}}\left(  \frac{-d\tau^{2}+d\rho^{2}}{\rho^{2}}%
+\Xi_{\circ}A_{\circ}^{2}d\Omega_{n-1}^{2}\right)  .
\end{equation}
This is in the BR form with the topological structure $AdS_{2}\times S^{N-2},$
where the radius of the sphere is $\sqrt{\Xi_{\circ}}A_{\circ}.$ It can be
shown that%
\begin{equation}
\Xi_{\circ}A_{\circ}^{2}=\left(  n-2\right)  \left(  \frac{8\beta^{2}%
Q^{2}-\left(  n-1\right)  \left(  n-2\right)  }{\left(  n-1\right)  \left(
n-2\right)  +8\beta^{2}Q^{2}}\right)
\end{equation}
which, in the limit of $\beta\rightarrow\infty,$ becomes
\begin{equation}
\underset{\beta\rightarrow\infty}{\lim}\Xi_{\circ}A_{\circ}^{2}=\left(
n-2\right)
\end{equation}
such that, the solution (81) becomes the BR type solution of EYMD theory (see
Eq. (39)). We set now $\Xi_{\circ}A_{\circ}^{2}=1$, to obtain a conformally
flat metric. This claims that
\begin{equation}
\left(  n-2\right)  \left(  \frac{8\beta^{2}Q^{2}-\left(  n-1\right)  \left(
n-2\right)  }{\left(  n-1\right)  \left(  n-2\right)  +8\beta^{2}Q^{2}%
}\right)  =1
\end{equation}
and consequently we find
\begin{align}
\beta^{2}  &  =\frac{\left(  n-1\right)  ^{2}\left(  n-2\right)  }%
{8Q^{2}\left(  n-3\right)  },\\
ds^{2}  &  =\frac{2Q^{2}}{\left(  n-1\right)  }\left(  \frac{-d\tau^{2}%
+d\rho^{2}}{\rho^{2}}+d\Omega_{3}^{2}\right)  .
\end{align}
This particular choice of $\beta$ casts the EYMBI metric into a conformally
flat form with the topology of $AdS_{2}\times S^{3}$

\subsubsection{$AdS_{2}\times S^{N-2}$ topology for $0<\alpha<1$}

As one may show, for $0<\alpha<1$ and $r_{+}=0,$ a similar transformation as
(40), here also leads to the line element
\begin{equation}
ds^{2}=\left(  \Xi\right)  ^{-\frac{1+\alpha^{2}}{1-\alpha^{2}}}\left(
\frac{1-\alpha^{2}}{1+\alpha^{2}}\right)  ^{-\frac{2}{1-\alpha^{2}}}%
\rho^{-\frac{2\alpha^{2}}{1-\alpha^{2}}}\left(  \frac{-d\tau^{2}+d\rho^{2}%
}{\rho^{2}}+\Xi A^{2}\left(  \frac{1-\alpha^{2}}{1+\alpha^{2}}\right)
^{2}d\Omega_{n-1}^{2}\right)  .
\end{equation}
Again we set $\Xi A^{2}\left(  \frac{1-\alpha^{2}}{1+\alpha^{2}}\right)
^{2}=1$ which gives the conformally flat line element%
\begin{equation}
ds^{2}=a\left(  \rho\right)  \left(  \frac{-d\tau^{2}+d\rho^{2}}{\rho^{2}%
}+d\Omega_{n-1}^{2}\right)  ,
\end{equation}
with%
\begin{equation}
a\left(  \rho\right)  =\left(  \Xi\right)  ^{-\frac{1+\alpha^{2}}{1-\alpha
^{2}}}\left(  \frac{1-\alpha^{2}}{1+\alpha^{2}}\right)  ^{-\frac{2}%
{1-\alpha^{2}}}\rho^{-\frac{2\alpha^{2}}{1-\alpha^{2}}}.
\end{equation}

\subsection{Linear Stability of the EYMBID black holes}

Similar to the proof given in Sec. ($II.B$), here also we study the stability
of the possible black holes in EYMBID theory which undergoes a linear
perturbation. Again we give a detailed study for the 5-dimensional black holes
which is extendible to any higher dimensions. Our perturbed metric is same as
we adapted in Eq. (46). The linearized field equations plus the extra term of
$R_{tr}$ are given now by%
\begin{gather}
R_{tr}:\frac{\left(  n-1\right)  }{2}\frac{\chi_{t}\left(  t,r\right)
h^{\prime}\left(  r\right)  }{h\left(  r\right)  }=\frac{4}{3}\partial_{r}%
\Phi\left(  r\right)  \partial_{t}\psi\left(  t,r\right) \\
\nabla_{\circ}^{2}\psi-\chi\nabla_{\circ}^{2}\Phi+\frac{1}{2}\left(
\Gamma-\chi\right)  _{r}\Phi^{\prime}f=-\frac{8}{\left(  n-1\right)  }%
\alpha^{2}\beta^{2}e^{\frac{4}{\left(  n-1\right)  }\alpha\Phi}\left(
\mathcal{L}\left(  X_{\circ}\right)  +4X_{\circ}^{2}\partial_{X_{\circ}}%
^{2}\mathcal{L}\left(  X_{\circ}\right)  \right)  \psi\\
R_{\theta\theta}:\left(  2-R_{\circ\theta\theta}\right)  \chi-\frac{1}%
{2}hh^{\prime}f\left(  \Gamma-\chi\right)  _{r}=\frac{16}{9}\alpha A^{2}%
\beta^{2}\left(  2X_{\circ}\partial_{X_{\circ}}\mathcal{L}\left(  X_{\circ
}\right)  -\mathcal{L}\left(  X_{\circ}\right)  \right)  \psi
\end{gather}
in which our conventions are as before. The first equation in this set implies
that%
\begin{equation}
\chi\left(  t,r\right)  =-\frac{4}{3\alpha}\psi\left(  t,r\right)
\end{equation}
which, after we make substitutions in the two latter equations and eliminating
the $\left(  \Gamma-\chi\right)  _{r}$ we find%
\begin{equation}
\nabla_{\circ}^{2}\psi\left(  t,r\right)  -U\left(  r\right)  \psi\left(
t,r\right)  =0
\end{equation}
where%
\begin{equation}
U\left(  r\right)  =\frac{8}{3}\beta^{2}e^{\frac{4}{3}\alpha\Phi}\left[
\mathcal{L}\left(  X_{\circ}\right)  -2X_{\circ}\partial_{X_{\circ}%
}\mathcal{L}\left(  X_{\circ}\right)  -\alpha^{2}\left(  \mathcal{L}\left(
X_{\circ}\right)  +4X_{\circ}^{2}\partial_{X_{\circ}}^{2}\mathcal{L}\left(
X_{\circ}\right)  \right)  \right]  .
\end{equation}
To get these results we have implicitly used the constraint (74) on $A$. Again
by imposing the same constraint , one can show that $U\left(  r\right)  $ is
positive definite. We follow the separation method to get
\begin{equation}
\psi\left(  t,r\right)  =e^{\pm\epsilon t}\zeta\left(  r\right)  ,\text{
\ \ }\nabla_{\circ}^{2}\zeta\left(  r\right)  -U_{eff}\left(  r\right)
\zeta\left(  r\right)  =0,\text{ \ \ }U_{eff}\left(  r\right)  =\left(
\frac{\epsilon^{2}}{f}+U\left(  r\right)  \right)  ,
\end{equation}
where $\epsilon$ is a constant. Here also the fact that $U_{eff}\left(
r\right)  >0$ can be justified which implies in turn that the system is
stable. For $\beta\rightarrow\infty$ this reduces to the case of EYMD black
hole solution whose stability was already verified before.

\section{Black holes in the BDYM theory}

In $N\left(  =n+1\right)  -$dimensions we write the Brans-Dicke-Yang-Mills
(BDYM) action as%
\begin{gather}
I=-\frac{1}{16\pi}\int\nolimits_{\mathcal{M}}d^{n+1}x\sqrt{-g}\left(  \phi
R-\frac{\omega}{\phi}\left(  \mathbf{\nabla}\phi\right)  ^{2}+\mathcal{L}%
_{m}\right)  -\frac{1}{8\pi}\int\nolimits_{\partial\mathcal{M}}d^{n}x\sqrt
{-h}K,\\
\mathcal{L}_{m}=-\mathbf{Tr}(F_{\lambda\sigma}^{\left(  a\right)  }F^{\left(
a\right)  \lambda\sigma}),\nonumber
\end{gather}
in which $\omega$ is the coupling constant, and $\phi$ stands for the BD
scalar field with the dimensions $G^{-1}$ ($G$ is the $N-$dimensional
Newtonian constant \cite{21}). Variation of the BDYM's action with respect to
the $g_{\mu\nu}$ gives
\begin{align}
\phi G_{\mu\nu}  &  =\frac{\omega}{\phi}\left(  \nabla_{\mu}\phi\nabla_{\nu
}\phi-\frac{1}{2}g_{\mu\nu}\left(  \mathbf{\nabla}\phi\right)  ^{2}\right)
+2\left(  \mathbf{Tr}\left(  F_{\mu\lambda}^{\left(  a\right)  }F_{\nu
}^{\left(  a\right)  \ \lambda}\right)  -\frac{1}{4}g_{\mu\nu}\mathbf{Tr}%
\left(  F_{\lambda\sigma}^{\left(  a\right)  }F^{\left(  a\right)
\lambda\sigma}\right)  \right)  +\\
&  \nabla_{\mu}\nabla_{\nu}\phi-g_{\mu\nu}\nabla^{2}\phi,\nonumber
\end{align}
while variation of the action with respect to the scalar field $\phi$ and the
gauge potentials $\mathbf{A}^{\left(  a\right)  }$ yields
\begin{equation}
\nabla^{2}\phi=-\frac{n-3}{2\left[  \left(  n-1\right)  \omega+n\right]
}\mathbf{Tr}\left(  F_{\lambda\sigma}^{\left(  a\right)  }F^{\left(  a\right)
\lambda\sigma}\right)  ,
\end{equation}
and%
\begin{equation}
\mathbf{d}\left(  ^{\star}\mathbf{F}^{\left(  a\right)  }\right)  +\frac
{1}{\sigma}C_{\left(  b\right)  \left(  c\right)  }^{\left(  a\right)
}\mathbf{A}^{\left(  b\right)  }\wedge^{\star}\mathbf{F}^{\left(  c\right)
}=0,
\end{equation}
respectively.

We follow now the routine process to transform BDYM action into the EYMD
action\cite{21}. For this purpose, one can use a conformal transformation
(variables with a caret $\hat{.}$ denote those in the Einstein frame)
\begin{equation}
\hat{g}_{\mu\nu}=\phi^{\frac{2}{n-1}}g_{\mu\nu}\text{ \ \ and \ \ }\hat{\Phi
}=\frac{\left(  n-3\right)  }{4\hat{\alpha}}\ln\phi.
\end{equation}
This transforms (97) into
\begin{equation}
\hat{I}=-\frac{1}{16\pi}\int\nolimits_{\mathcal{M}}d^{n+1}x\sqrt{-\hat{g}%
}\left(  \hat{R}-\frac{4}{n-1}\left(  \mathbf{\hat{\nabla}}\hat{\Phi}\right)
^{2}-e^{-4\hat{\alpha}\hat{\Phi}/\left(  n-1\right)  }\mathbf{Tr}\left(
\hat{F}_{\lambda\sigma}^{\left(  a\right)  }\hat{F}^{\left(  a\right)
\lambda\sigma}\right)  \right)  -\frac{1}{8\pi}\int\nolimits_{\partial
\mathcal{M}}d^{n}x\sqrt{-\hat{h}}\hat{K},
\end{equation}
where
\begin{equation}
\hat{\alpha}=\frac{n-3}{2\sqrt{\left(  n-1\right)  \omega+n}}.
\end{equation}
This transformed action is similar to the EYMD action given by (1). Variation
of this action with respect to the $\hat{g}_{\mu\nu},$ $\hat{\Phi}$ and
$\mathbf{\hat{A}}^{\left(  a\right)  }$ gives%
\begin{gather}
\hat{R}_{\mu\nu}=\frac{4}{n-1}\mathbf{\hat{\partial}}_{\mu}\Phi\hat{\partial
}_{\nu}\Phi+2e^{-4\hat{\alpha}\hat{\Phi}/\left(  n-1\right)  }\left[
\mathbf{Tr}\left(  \hat{F}_{\mu\lambda}^{\left(  a\right)  }\hat{F}_{\nu
}^{\left(  a\right)  \ \lambda}\right)  -\frac{1}{2\left(  n-1\right)
}\mathbf{Tr}\left(  \hat{F}_{\lambda\sigma}^{\left(  a\right)  }\hat
{F}^{\left(  a\right)  \lambda\sigma}\right)  \hat{g}_{\mu\nu}\right]  ,\\
\hat{\nabla}^{2}\Phi=-\frac{1}{2}\hat{\alpha}e^{-4\hat{\alpha}\hat{\Phi
}/\left(  n-1\right)  }\mathbf{Tr}(\hat{F}_{\lambda\sigma}^{\left(  a\right)
}\hat{F}^{\left(  a\right)  \lambda\sigma}),
\end{gather}

\begin{equation}
\mathbf{d}\left(  e^{-4\hat{\alpha}\hat{\Phi}/\left(  n-1\right)  \star
}\mathbf{\hat{F}}^{\left(  a\right)  }\right)  +\frac{1}{\sigma}C_{\left(
b\right)  \left(  c\right)  }^{\left(  a\right)  }e^{-4\hat{\alpha}\hat{\Phi
}/\left(  n-1\right)  }\mathbf{\hat{A}}^{\left(  b\right)  }\wedge^{\star
}\mathbf{\hat{F}}^{\left(  c\right)  }=0.
\end{equation}
It is not difficult to conclude that, if we find a solution to the latter
equations, by an inverse transformation, we can find the solutions of the
related equations of the BDYM theory. In other words if $\left(  \hat{g}%
_{\mu\nu},\Phi,\mathbf{\hat{F}}^{\left(  a\right)  }\right)  $ is a solution
of the latter equations, then
\begin{equation}
\left(  g_{\mu\nu},\phi,\mathbf{F}^{\left(  a\right)  }\right)  =\left(
\exp\left(  -\frac{8\hat{\alpha}}{\left(  n-1\right)  \left(  n-3\right)
}\hat{\Phi}\right)  \hat{g}_{\mu\nu},\exp\left(  \frac{4\hat{\alpha}}{\left(
n-3\right)  }\hat{\Phi}\right)  ,\mathbf{\hat{F}}^{\left(  a\right)  }\right)
\end{equation}
is a solution of (98-100) and vice versa.

One may call $\left(  g_{\mu\nu},\phi,\mathbf{F}^{\left(  a\right)  }\right)
,$ the reference solution and $\left(  \hat{g}_{\mu\nu},\hat{\Phi
},\mathbf{\hat{F}}^{\left(  a\right)  }\right)  $ the target solution. Hence
our solution in EYMD would be the target solution i.e.%
\begin{equation}
d\hat{s}^{2}=-\hat{f}\left(  r\right)  dt^{2}+\frac{dr^{2}}{\hat{f}\left(
r\right)  }+\hat{h}\left(  r\right)  ^{2}d\Omega_{n-1}^{2},
\end{equation}
where%
\begin{align}
\hat{f}\left(  r\right)   &  =\hat{\Xi}\left(  1-\left(  \frac{\hat{r}_{+}}%
{r}\right)  ^{\frac{\left(  n-2\right)  \hat{\alpha}^{2}+1}{\hat{\alpha}%
^{2}+1}}\right)  r^{\frac{2}{\hat{\alpha}^{2}+1}},\text{ \ \ }\hat{h}\left(
r\right)  =\hat{A}e^{-2\hat{\alpha}\hat{\Phi}/\left(  n-1\right)  },\\
\hat{\Xi}  &  =\frac{\left(  n-2\right)  }{\left(  \left(  n-2\right)
\hat{\alpha}^{2}+1\right)  \hat{Q}^{2}},\text{ \ \ }\hat{\Phi}=-\frac{\left(
n-1\right)  }{2}\frac{\hat{\alpha}\ln r}{\hat{\alpha}^{2}+1},\text{ \ \ }%
\hat{A}^{2}=\hat{Q}^{2}\left(  \hat{\alpha}^{2}+1\right)  ,\nonumber\\
\hat{r}_{+}  &  =\left(  \frac{4\left(  \hat{\alpha}^{2}+1\right)  \hat
{M}_{QL}}{\left(  n-1\right)  \hat{\Xi}\hat{\alpha}^{2}\hat{A}^{n-1}}\right)
.\nonumber
\end{align}
Our reference solution would read now%
\begin{equation}
ds^{2}=-f\left(  r\right)  dt^{2}+\frac{dr^{2}}{f\left(  r\right)  }+h\left(
r\right)  ^{2}d\Omega_{n-1}^{2},
\end{equation}
in which%
\begin{align}
f\left(  r\right)   &  =\hat{\Xi}\left(  1-\left(  \frac{\hat{r}_{+}}%
{r}\right)  ^{\frac{\left(  n-2\right)  \hat{\alpha}^{2}+1}{\hat{\alpha}%
^{2}+1}}\right)  r^{\frac{2\left(  n-3\right)  +4\hat{\alpha}^{2}}{\left(
n-3\right)  \left(  \hat{\alpha}^{2}+1\right)  }},\text{ \ \ }h\left(
r\right)  =\hat{A}e^{-\frac{2\hat{\alpha}\hat{\Phi}\left(  n+1\right)
}{\left(  n-1\right)  \left(  n-3\right)  }}=\hat{A}r^{\frac{\hat{\alpha}%
^{2}\left(  n+1\right)  }{\left(  \hat{\alpha}^{2}+1\right)  \left(
n-3\right)  }},\\
\phi &  =r^{\frac{-2\left(  n-1\right)  \hat{\alpha}^{2}}{\left(  n-3\right)
\left(  \hat{\alpha}^{2}+1\right)  }},\text{ and }\mathbf{F}^{\left(
a\right)  }=\mathbf{\hat{F}}^{\left(  a\right)  }=\mathbf{d\hat{A}}^{\left(
a\right)  }+\frac{1}{2\sigma}C_{\left(  b\right)  \left(  c\right)  }^{\left(
a\right)  }\mathbf{\hat{A}}^{\left(  b\right)  }\wedge\mathbf{\hat{A}%
}^{\left(  c\right)  }%
\end{align}
where the YM potential is same as (9) with the new charge $\hat{Q}.$ Herein
one can find the Hawking temperature of the BDYM-black hole at the event
horizon as%
\begin{equation}
T_{H}=\frac{\hat{\Xi}\left[  \left(  n-2\right)  \hat{\alpha}^{2}+1\right]
}{4\pi\left(  \hat{\alpha}^{2}+1\right)  }\left(  \hat{r}_{+}\right)
^{\left(  -\frac{\left(  n-3\right)  \left(  \hat{\alpha}^{2}-1\right)
-4\hat{\alpha}^{2}}{\left(  \hat{\alpha}^{2}+1\right)  \left(  n-3\right)
}\right)  }%
\end{equation}
where $\hat{r}_{+}$ is the radius of the event horizon.

\section{Conclusion}

A simple class of spherically symmetric solutions to the EYMD equations is
obtained in any dimensions. Magnetic type Wu-Yang ansatz played a crucial role
in extending the solution to N-dimension. For the non-zero dilaton the space
time possesses singularity, representing a non-asymptotically flat black hole
solution expressed in terms of the quasilocal mass. Particular case of a
linear dilatonic black hole is singled out as a specific case. Hawking
temperature for all cases has been computed which are distinct from the EMD
temperatures \cite{22}. Stability against linear perturbations for these
dilatonic metrics is proved. It has been shown that the extremal limit in the
vanishing dilaton, results in the higher dimensional BR space times for the YM
field. With the common topology of $AdS_{2}\times S^{N-2}$ for both theories,
while the radius of $S^{N-2}$ for the Maxwell case is $\left(  N-3\right)  ,$
it becomes $\left(  N-3\right)  ^{1/2}$ in the YM case. As a final
contribution in the paper we apply a conformal transformation to derive black
hole solutions in the Brans-Dicke-YM theory. It is our belief that these YMBR
metrics, beside the dilatonic ones, will be useful in the string/supergravity
theory as much as the EMBR metrics are.

\begin{acknowledgement}
We thank the anonymous referee for valuable and constructive suggestions.
\end{acknowledgement}

\section{Appendix A}

\bigskip We work on a group of proper rotations in $\left(  N-1\right)
-$dimensions, $SO(N-1),$ which forms a $\frac{\left(  N-1\right)  \left(
N-2\right)  }{2}\left(  \text{i.e., }\dbinom{N-1}{2}\right)  -$parameter Lie
group whose infinitesimal generators are given by:%
\begin{align}
L_{1}  &  =x_{2}\partial_{x_{1}}-x_{1}\partial_{x_{2}}\tag{A-1}\\
L_{2}  &  =x_{3}\partial_{x_{1}}-x_{1}\partial_{x_{3}}\nonumber\\
L_{3}  &  =x_{3}\partial_{x_{2}}-x_{2}\partial_{x_{3}}\nonumber\\
L_{4}  &  =x_{4}\partial_{x_{1}}-x_{1}\partial_{x_{4}}\nonumber\\
L_{5}  &  =x_{4}\partial_{x_{2}}-x_{2}\partial_{x_{4}}\nonumber\\
L_{6}  &  =x_{4}\partial_{x_{3}}-x_{3}\partial_{x_{4}}\nonumber\\
&  ....\nonumber
\end{align}

These operators satisfy commutation relations of the form%
\begin{equation}
\left[  L_{i},L_{j}\right]  =C_{\left(  i\right)  \left(  j\right)  }^{\left(
k\right)  }L_{k}, \tag{A-2}%
\end{equation}
where the $C_{\left(  i\right)  \left(  j\right)  }^{\left(  k\right)  }$ are
the structure constants. As an example we check
\begin{align}
\left[  L_{1},L_{2}\right]   &  =C_{\left(  1\right)  \left(  2\right)
}^{\left(  3\right)  }L_{3}=L_{3},\tag{A-3}\\
&  \rightarrow C_{\left(  1\right)  \left(  2\right)  }^{\left(  3\right)
}=1.\nonumber
\end{align}

This can be done for all other combinations and the only 24 non zero terms
are:%
\begin{equation}%
\begin{tabular}
[c]{ll}%
$C_{\left(  2\right)  \left(  3\right)  }^{\left(  1\right)  }=C_{\left(
4\right)  \left(  5\right)  }^{\left(  1\right)  }=-C_{\left(  3\right)
\left(  2\right)  }^{\left(  1\right)  }=-C_{\left(  5\right)  \left(
4\right)  }^{\left(  1\right)  }=$ & $1$\\
$C_{\left(  3\right)  \left(  1\right)  }^{\left(  2\right)  }=C_{\left(
4\right)  \left(  6\right)  }^{\left(  2\right)  }=-C_{\left(  1\right)
\left(  3\right)  }^{\left(  2\right)  }=-C_{\left(  6\right)  \left(
4\right)  }^{\left(  2\right)  }=$ & $1$\\
$C_{\left(  1\right)  \left(  2\right)  }^{\left(  3\right)  }=C_{\left(
5\right)  \left(  6\right)  }^{\left(  3\right)  }=-C_{\left(  2\right)
\left(  1\right)  }^{\left(  3\right)  }=-C_{\left(  6\right)  \left(
5\right)  }^{\left(  3\right)  }=$ & $1$\\
$C_{\left(  5\right)  \left(  1\right)  }^{\left(  4\right)  }=C_{\left(
6\right)  \left(  2\right)  }^{\left(  4\right)  }=-C_{\left(  1\right)
\left(  5\right)  }^{\left(  4\right)  }=-C_{\left(  2\right)  \left(
6\right)  }^{\left(  4\right)  }=$ & $1$\\
$C_{\left(  1\right)  \left(  4\right)  }^{\left(  5\right)  }=C_{\left(
6\right)  \left(  3\right)  }^{\left(  5\right)  }=-C_{\left(  4\right)
\left(  1\right)  }^{\left(  5\right)  }=-C_{\left(  3\right)  \left(
6\right)  }^{\left(  5\right)  }=$ & $1$\\
$C_{\left(  2\right)  \left(  4\right)  }^{\left(  6\right)  }=C_{\left(
3\right)  \left(  5\right)  }^{\left(  6\right)  }=-C_{\left(  4\right)
\left(  2\right)  }^{\left(  6\right)  }=-C_{\left(  5\right)  \left(
3\right)  }^{\left(  6\right)  }=$ & $1$%
\end{tabular}
\ \ \tag{A-4}%
\end{equation}

By a similar, routine procedure we can obtain the coefficients in any higher
dimensions. For $N=6$, for example, we have 40 non-zero coefficients, which we
shall not elaborate.

\end{document}